\documentclass{iau}

\usepackage{amsmath}
\usepackage{graphicx}
\usepackage{multirow}

\begin{document}

\lefttitle{Peter Frinchaboy \& MSE Team}
\righttitle{Stellar Populations with the Maunakea Spectroscopic Explorer}

\jnlPage{1}{7}
\jnlDoiYr{2025}
\doival{10.1017/xxxxx}
\volno{395}
\pubYr{2025}
\journaltitle{Stellar populations in the Milky Way and beyond}

\aopheadtitle{Proceedings of the IAU Symposium}
\editors{J. Mel\'endez,  C. Chiappini, R. Schiavon \& M. Trevisan, eds.}

\title{The Future of Stellar Populations with the Maunakea Spectroscopic Explorer (MSE)}

\author{Peter Frinchaboy$^{1,2}$, Andy Sheinis$^2$, Sam Barden$^2$, Viraja Khatu$^2$}
\affiliation{$^1$Department of Physics \& Astronomy, Texas Christian University, Fort Worth, TX 76129, USA}
\affiliation{$^2$Canada-France-Hawaii Telescope, 65-1238 Mamalahoa Highway, Kamuela, HI 96743, USA}

\begin{abstract}
The Maunakea Spectroscopic Explorer (MSE) ia a massively multiplexed spectroscopic survey facility that is proposed to replace the Canada-France-Hawai'i-Telescope in the 2040s.  Since 2019, due to the uncertainty for new facilities on Maunakea, the project has been focused on new technology enabling greater capabilities beyond the concept design reviewed facility plan.  Enhanced fiber density, and thereby survey speed, is made possible by using a new quad mirror (QM) 11.5-meter telescope design with 18,000+ fibers and a 1.5 square degree field-of-view.  The MSE spectrographs will be moderate-resolution (360 nm through H-band at R=7,000) and high-resolution (R=40,000). MSE’s baseline NIR capabilities will enable studies of highly-reddened regions in the Local Group, unlike other proposed next generation facilities. The MSE large-scale survey instrument suite will enable the equivalent to a full SDSS Legacy Survey every several weeks.  This work presents the current status of the project after the Fall 2024 MSE science Workshop.
\end{abstract}


\maketitle
\vskip-0.16in
\section{Introduction}
The Maunakea Spectroscopic Explorer (MSE; mse.cfht.hawaii.edu) project has eight science working groups, five of which are relevant for local group studies (Exoplanets and Stellar Astrophysics, Milky Way and Resolved Stellar Population, Chemical Nucleosynthesis, Astrophysical Tests of Dark Matter, and Time Domain Astronomy and Transients).  This work presents the discussions related to these working groups and new technology and design work from the 2024 MSE Virtual Meeting held between October 28th and November 8th, 2024.\\

MSE's goal is to carry out the ultimate spectroscopic follow-up of the Gaia mission and is critical to our understanding of the faint and distant regimes of the Galaxy and Local Group. Uniquely, MSE will conduct in-situ chemodynamical analysis of individual stars in all Galactic components, searching for inter-relationships between them and for departures from equilibrium. MSE will bring about an entirely new era for nearby dwarf galaxy studies, enabling accurate chemo-dynamical measurements to be performed efficiently across the full range of dwarf galaxy luminosities ($10^{3-7}$ L$_{\odot}$), and providing spectra for at least an order of magnitude more stars in each system. MSE will also provide a comprehensive understanding of the chemodynamics of M31 and M33.\\

This work comes after the workshop dedicated to updating the community on possible new technology designs, lessons learned from other large spectroscopic survey projects (SDSS, 4MOST, LAMOST, GALAH, DESI, PFS), and starting to evaluate should the existing MSE science requirements be updated based on new science since the 2019 science case (The MSE Science Team, 2019) and due to new science discoveries and the MSE project's delayed schedule. \\

\clearpage
\section{MSE New Design work}
MSE has completed the concept design review, however due to delays with Maunakea, we have reopened consideration of the design of the telescope, focal plane, and spectrographs, which has led to the Quad-Mirror design (that enables a five fold increase in fiber density; Barden et al 2024; Figure 1 {\em left}). This new design leads to a number of key improvements, (1) allows for a larger primary, primarily limited due to size of the largest transmissive optical element, (2) move the fiber focal plane from prime-focus to Nasmyth focus enabling a physically larger focal plane, (3) the Nasmyth focal plane enables packing 18,000+ fibers, and (4) the Nasmyth focus allows the possibility of significantly shorter fibers for spectrographs that could be placed on the nasmyth platform, which could allow $\sim 1$ magnitude improvement in near-UV throughput.  \\

Due to difficulty in producing the needed high-resolution spectrographs and to take advantage to the potential of increased blue throughput, MSE work has been done to develop and test the concept of pupil and wavelength splitting (Barden 2024; Figure 1 {\em right}).  Through a NSF ATI grant (NSF-2307501), Sam Barden has developed and tested a prototype pupil and wavelength splitting system. The potential of these systems are two-fold, (1) wavelength splitting at the fiber level enables 
(2) pupil splitting enables the each fiber to "see" a smaller pupil enabling the use of smaller and cheaper spectrographs, but also requiring more of them.  
Pupil and wavelength splitting can be implemented separately or on combination, and we expect MSE and other future projects to consider implementing these technologies.\\ 

The current baseline MSE QM-based instrument/focal plane configuration includes $\sim 14,000+$ fibers feeding 5-6 channel medium-resolution (MR) spectrographs covering the full 0.4-1.3$\mu$m and $H$-band (1.5-1.7$\mu$m) with $R \sim 5000$ and $4000+$ fibers feeding high- resolution (HR) spectrographs with 3 channels (401-471 nm @ R = 40k, 471-489 nm @ R = 40k, and 625-674 nm @ R = 20k).  At the workshop, new possible MSE MR and LR spectrograph designs were presented, including a folded Schmidt camera design, that would enable less optical obstruction and smaller/easier detector mounting (Will Saunders, private communication).  

%\clearpage
\begin{figure}[b!]
  \centerline{\vbox to 1pc{\hbox to 10pc{}}}
  \includegraphics[scale=.23]{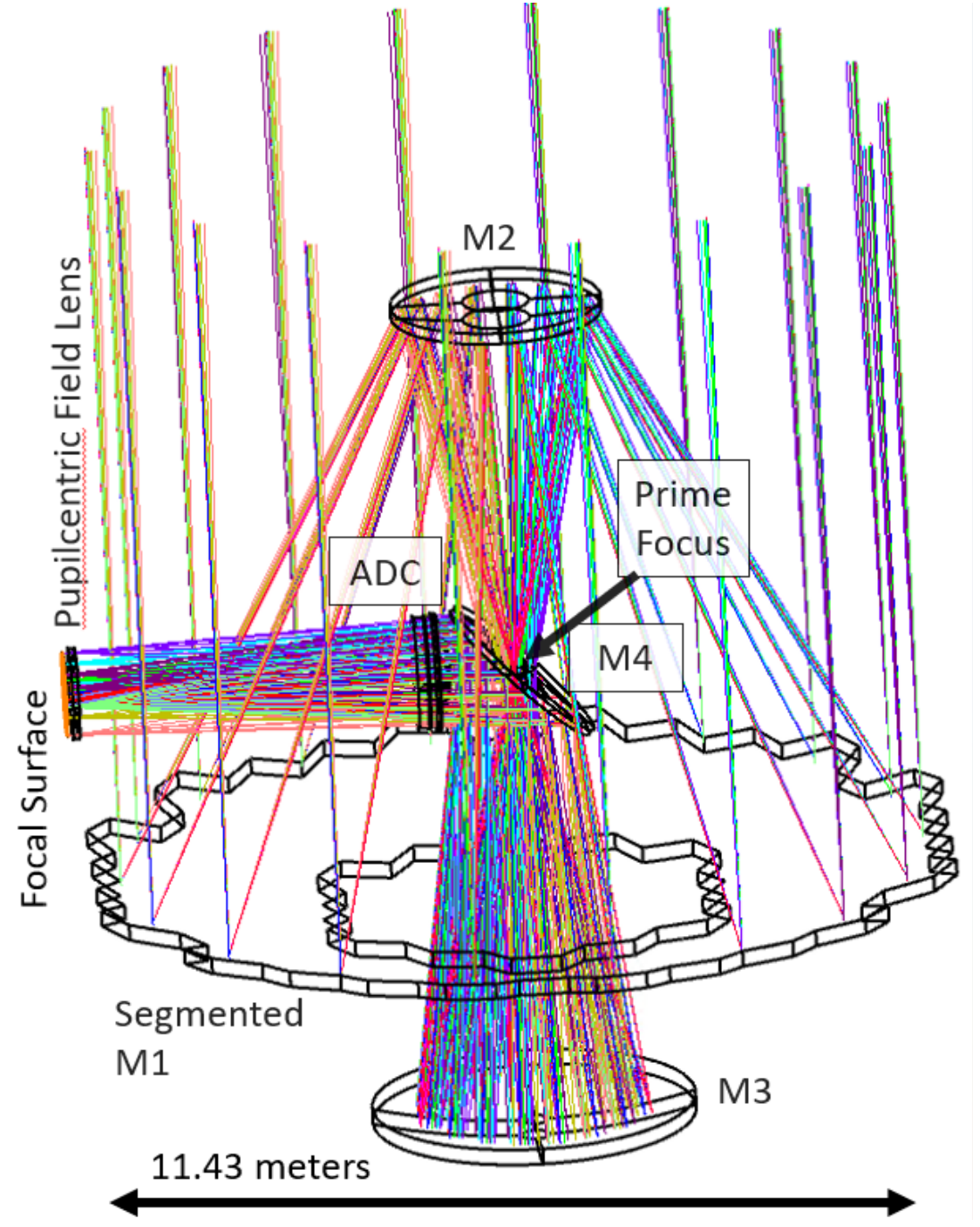}
  \includegraphics[scale=.39]{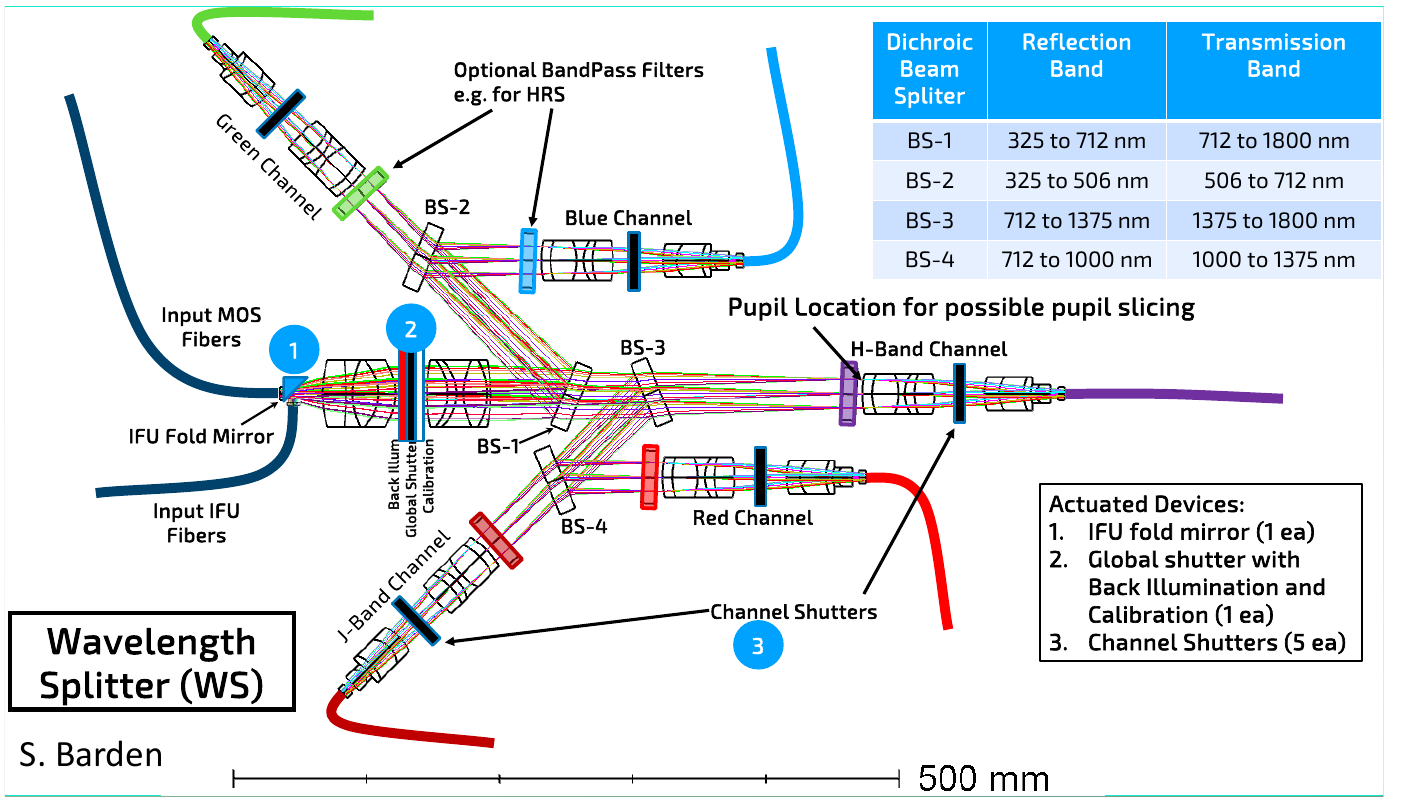}
  \caption{(Left) The MSE Quad Mirror (QM) design that is the now current proposed baseline for the MSE telescope, designed by Sam Barden (Barden 2024).  This design could be scaled up, but is currently constrained considering working within the existing CFHT footprint and by realistic assessment of the producibility of largest optical element. (Right) The new proposed pupil and wavelength-splitting technology for MSE, which will enable smaller and cheaper, but more, spectrographs to be built to meet the MSE science requirements.  This viability of this technology is being evaluated by Sam Barden at CFHT supported by an NSF ATI grant (AST-2307501; Barden et al. 2024).}
  \label{MSE}
\end{figure}

\clearpage 
New potential fiber positioner designs, with significantly larger positioner patrol range were presented (Roloef de Jong, private communication) that are currently starting prototype development.  This or similar technology could allow MSE to better populate the high resolution (HR) fibers, while also reduce the total number of HR spectrographs needed per wavelength channel. 

%\clearpage
\section{MSE Workshop 2024 - Stellar Populations Science Implications}

\noindent {\bf The key stellar populations science questions discussed and their implications on MSE science requirement/design work:}\\
\begin{itemize}
\item What is the optimal High-Resolution (HR) fiber density, considering potential new positioner technologies, target densities, and the physically larger QM focal plane.\\
\item Evaluate trade-off between the number of HR fibers in the focal plane (currently $\sim$ 1/3 of the fibers would be HR to allow coverage over the full focal plane) and how many HR spectrograph channels/wavelength regions (currently only 3 HR channels) should be covered to produce the groundbreaking science in the 2040s.\\
\item Determine what HR resolution is {\em needed} ($R \sim$ 20,000--80,000) to reliably measure key enrichment tracing abundances (examples included Fe, Zn, Mn, U, Th, C, N) to the desired precision.  Particularly taking into account that at different metallicities will require different resolutions, e.g., line blanketing in metal-rich stars ([Fe/H] $\ge 0$) may actually require higher resolution than metal-poor ([Fe/H] $< 2$) stars. \\
\item For the medium resolution (MR) spectrographs is $R \sim 5000$ sufficient for stellar population science needs?  Would one channel need to be higher resolution to enable more science, e.g., higher precision velocities or more elements ([$\alpha$/Fe], can other elements be measured, carbon, nitrogen, oxygen from molecules with NIR at $R \sim 5000$)?\\
\item What new science is enabled by taking advantage of the Maunakea site through using the Near-UV (310-390 nm) sky?  \\
\item Is the science enabled by adding high resolution Near-UV and/or $H$-band science worth the additional costs (UV optimized spectrograph, cryogenic NIR spectrograph), and potentially increased detector cost?  Does wavelength splitting work well enough to enable these options?  
\end{itemize}

\end{document}